\documentclass[12pt]{article}
\usepackage{a4wide,amssymb,cite}
\pdfoutput=1

\usepackage{a4wide,amssymb,graphicx}
\usepackage{epsfig}
\usepackage[usenames,dvipsnames]{color}
\usepackage{slashed}
\usepackage{amssymb,cite,graphicx}
\usepackage{slashed}
\usepackage{amsmath,bm,bbm}
\usepackage{amsfonts}
\usepackage[titletoc,title]{appendix}
\usepackage[small]{caption}
\usepackage[margin=1in]{geometry}
\usepackage[multiple]{footmisc}
\usepackage{mathtools}
\usepackage{slashed}
\usepackage[nottoc]{tocbibind}
\usepackage{xcolor}

\newcommand{\be}{\begin{equation}}
\newcommand{\ee}{\end{equation}}
\newcommand{\bea}{\begin{eqnarray}}
\newcommand{\eea}{\end{eqnarray}}

\def\circa#1{\,\raise.3ex\hbox{$#1$\kern-.75em\lower1ex\hbox{$\sim$}}\,}

\begin{document}

\begin{titlepage}
%
%


%

\begin{centering}
\vspace{1cm}
{\Large {\bf Higgs Inflation at the Pole}} \\

\vspace{1.5cm}

{\bf Simon Cl\'ery$^{1,\ddagger}$, Hyun Min Lee$^{2,\dagger}$ and Adriana G. Menkara$^{2,\sharp}$ }
\\
\vspace{.5cm}

{\it $^1$Universit\'e Paris-Saclay, CNRS/IN2P3, IJCLab, 91405 Orsay, France.} 
\\
{\it $^2$Department of Physics, Chung-Ang University, Seoul 06974, Korea.}

\vspace{.5cm}


\end{centering}
\vspace{2cm}

\begin{abstract}
\noindent
We propose a novel possibility for Higgs inflation where the perturbative unitarity below the Planck scale is ensured by construction and the successful predictions for inflation are accommodated. The conformal gravity coupling for the Higgs field leads to the proximity of the effective Planck mass to zero in the Jordan frame during inflation, corresponding to a pole in the Higgs kinetic term in the Einstein frame. Requiring the Higgs potential to vanish at the conformal pole in the effective theory in the Jordan frame, we make a robust prediction of the successful Higgs inflation. For a successful Higgs inflation at the pole, we take the running quartic coupling for the Higgs field to be small enough at the inflation scale, being consistent with the low-energy data, but we need a nontrivial extension of the SM with extra scalar or gauge fields in order to keep the running Higgs quartic coupling small during inflation. Performing the perturbative analysis of reheating with the known couplings of the SM particles to the Higgs boson, we show that a concrete realization of the Higgs pole inflation can be pinned down by the reheating processes with a general equation of state for the Higgs inflaton.  We illustrate some extensions of the simple Higgs pole inflation to the general pole expansions, the running Higgs quartic coupling in the Standard Model and its extension with a singlet scalar field, a supergravity embedding of the Higgs pole inflation.

\end{abstract}

\vspace{3cm}

\begin{flushleft} 
${}^{\ddagger}$Email: simon.clery@ijclab.in2p3.fr \\
$^\dagger$Email: hminlee@cau.ac.kr \\
$^\sharp$Email: amenkara@cau.ac.kr 
\end{flushleft}

\end{titlepage}

\section{Introduction}

Higgs inflation \cite{higgsinf} has drawn a lot of attention due to the discovery of the Higgs boson and its minimality with one more parameter beyond the Standard Model, the so called non-minimal coupling. It is very interesting in general to make a connection between the low-energy phenomena such as Higgs data, dark matter, and the high-energy physics in the early universe such as inflation.  However, the original proposal for Higgs inflation itself is lacking a predictive power because a large non-minimal coupling gives rise to unitarity violation below the Planck scale \cite{problem}. 
Thus, linear sigma-model constructions have been proposed for unitarizing the Higgs inflation \cite{unitarity}, and they are found to be dual to the effective theory beyond the Einstein gravity with $R^2$ term and maintain similar inflationary predictions as in the original Higgs inflation \cite{ema,general,HiggsR2susy,HiggsR2}.

In this article, we pursue an alternative route for Higgs inflation in the effective theory without a large non-minimal coupling. We take an expansion of the Higgs field around the pole for which the effective Planck mass in the Jordan frame vanishes for a conformal coupling to the Higgs or equivalently the Higgs kinetic term in the Einstein frame diverges. As a result, we show that a successful slow-roll inflation with the Higgs relies on the condition that the effective Higgs potential in the Jordan frame vanishes at the pole as well.  We dub this possibility the Higgs pole inflation. 
We take the Higgs potential with arbitrary single power dominance in the Einstein frame and obtain the inflationary predictions of the Higgs pole inflation as in Starobinsky inflation \cite{Starobinsky:1980te}. For a successful Higgs pole inflation, we derive the bounds on the other renormalizable and non-renormalizable terms in the general Higgs potential in the effective theory. In particular, it is necessary to impose a small running quartic coupling for the Higgs field during inflation, which is consistent with the low-energy data but calls for a nontrivial extension of the SM. We remark that the Higgs pole inflation belongs to a class of inflation models containing a pole of the kinetic term, called the $\alpha$-attractors in supergravity \cite{alphatt}.

For reheating, we take into account the general equation of state of the Higgs inflaton and the perturbative processes such as decay and scattering of the Higgs inflaton into the SM particles. We also show the effects of preheating due to the inflaton oscillations in the region where the perturbative processes are blocked kinematically.  It is crucial to notice that the known couplings of the SM particles to the Higgs boson allow for a reliable analysis of reheating in our work. We propose several extensions of the Higgs pole inflation by the general pole expansions of the effective Lagrangian and the running Higgs quartic coupling in the SM and its extensions with a singlet scalar field. In this case, it is important to make sure that two-loop and higher-loop beta function coefficients of the Higgs quartic coupling can be sufficiently small near the inflation scale in the extensions of the SM such that a small Higgs quartic coupling is maintained during inflation. Furthermore, we embed the Higgs pole inflation in supergravity where a general superpotential for two Higgs doublets generates an F-term potential for the Higgs pole inflation while the effective Higgs quartic coupling is dynamically relaxed to a small value along the D-flat direction during inflation. 
  
The paper is organized as follows.
We first describe the setup for the effective theory for Higgs inflation and the minimal pole inflation with the Higgs boson. Then, we discuss the inflationary dynamics and the bounds on the general Higgs potential. Next, we consider a general inflaton condensate for the anharmonic Higgs potential during reheating and analyze perturbative reheating and preheating in our model. We also present the extensions of the Higgs pole inflation by the general pole expansions, the running Higgs quartic coupling, and the embedding into the supergravity models. Finally, conclusions are drawn.

\section{The setup}

We consider the effective theory for Higgs inflation below the Planck scale. We take the SM Higgs to be conformally coupled to gravity in the leading order, but the conformal symmetry is broken by the mass term and the higher order interactions in either the non-minimal coupling or the Higgs potential. Then, the general Lagrangian for the Higgs inflation is given by
\bea
\frac{{\cal L}_J}{\sqrt{-g_J}} = -\frac{1}{2}M^2_P\, \Omega(H) R(g_J) + |D_\mu H|^2 -V_J(H) \label{LJ}
\eea
where the non-minimal coupling function and the effective Higgs potential are taken to
\bea
\Omega(H) &=& 1-\frac{1}{3M^2_P}|H|^2+ \sum_{n=2}^\infty\frac{b_n}{\Lambda^{2n-4}} \, |H|^{2n}, \\
V_J(H) &=& \mu^2_H |H|^2 +\lambda_H |H|^4 + \sum_{n=3}^\infty \frac{c_n}{\Lambda^{2n-4}} \, |H|^{2n}.  \label{HEFT}
\eea
Here, $\Lambda$ is the cutoff scale, and $b_n, c_n$ are dimensionless parameters.

Suppose that all $b_n$'s in the non-minimal coupling function $\Omega(H)$ are set to zero.
Then, after making a Weyl transformation of the metric by $g_{J,\mu\nu}=g_{E,\mu\nu}/\Omega$ with $\Omega=1-\frac{1}{3M^2_P}|H|^2$, we obtain the Einstein-frame Lagrangian as follows,
\bea
\frac{{\cal L}_E}{\sqrt{-g_E}} &=&-\frac{1}{2} M^2_P R(g_E) + \frac{|D_\mu H|^2}{\big(1-\frac{1}{3M^2_P}|H|^2\big)^2} \nonumber \\
&&-\frac{1}{3M^2_P\big(1-\frac{1}{3M^2_P}|H|^2\big)^2}\bigg(|H|^2 |D_\mu H|^2-\frac{1}{4}\partial_\mu |H|^2 \partial^\mu |H|^2\bigg) -V_E(H) \label{LE}
\eea
where the Einstein-frame Higgs potential becomes
\bea
V_E(H)= \frac{V_J(H)}{\big(1-\frac{1}{3M^2_P}|H|^2\big)^2}.
\eea

In the unitary gauge where the Higgs doublet is taken to $H^T=(0,h)^T/\sqrt{2}$, the first two terms in the second line of eq.~(\ref{LE}) are cancelled. Then, we get the Einstein-frame Lagrangian for the Higgs boson $h$ as
\bea
\frac{{\cal L}_E}{\sqrt{-g_E}} =-\frac{1}{2} M^2_P R +\frac{1}{2}\,\frac{(\partial_\mu h)^2}{\big(1-\frac{1}{6M^2_P}h^2\big)^2} - \frac{V_J\big(\frac{1}{\sqrt{2}}h\big)}{\big(1-\frac{1}{6M^2_P}h^2\big)^2}. \label{Linf}
\eea
Then, making the Higgs kinetic term canonically normalized by
\bea
h=\sqrt{6}M_P \tanh\Big(\frac{\phi}{\sqrt{6}M_P}\Big), \label{can}
\eea
we rewrite the Einstein-frame Lagrangian in eq.~(\ref{Linf}) as
\bea
\frac{{\cal L}_E}{\sqrt{-g_E}} =-\frac{1}{2} M^2_P R + \frac{1}{2}(\partial_\mu\phi)^2 - V_E(\phi),
\eea
with 
\bea
V_E(\phi)= \cosh^4\Big(\frac{\phi}{\sqrt{6}M_P}\Big) V_J\bigg(\sqrt{3}\tanh\Big(\frac{\phi}{\sqrt{6}M_P}\Big)\bigg). \label{Einpot}
\eea

As a result, the success of a slow-roll Higgs inflation depends on the form of the Jordan frame potential. 
We take the minimal form of the effective potential for the Higgs pole inflation in the Jordan frame and generalize it in the later sections. 

As a working example for inflation, we take a simple form of the Jordan-frame Higgs potential in eq.~(\ref{HEFT}), as follows,
\bea
V_J(H) = c_m \Lambda^{4-2m} |H|^{2m} \bigg(1-\frac{1}{3M^2_P}|H|^2\bigg)^2,  \label{poleJ}
\eea 
which leads to the the Einstein frame potential as
\bea
V_E(H)=c_m \Lambda^{4-2m} |H|^{2m}. \label{poleE}
\eea
This case corresponds to a single-power dominance of the Higgs potential in the Einstein frame.
Then, in terms of the canonical Higgs field $\phi$ given in eq.~(\ref{can}), the Einstein-frame potential in eq.~(\ref{Einpot}) becomes
\bea
V_E(\phi)&=& \frac{c_m}{2^m}  \Lambda^{4-2m} h^{2m} \nonumber \\
&=& 3^m c_m\Lambda^{4-2m}M^{2m}_P \bigg[  \tanh\Big(\frac{\phi}{\sqrt{6}M_P}\Big)\bigg]^{2m}. \label{inflatonpot}
\eea 
Therefore, a slow-roll inflation is possible at large $|\phi|$ or $|h|\sim \sqrt{6} M_P$ near the pole of the general Higgs kinetic term in eq.~(\ref{Linf}). We dub this possibility the Higgs pole inflation.

\section{Inflationary dynamics and Higgs parameters}

 In this section, we discuss the inflationary predictions in the Higgs pole inflation for the Einstein-frame Higgs potential with a single power dominance. Then, we show the bounds on the Higgs potential terms for the consistent Higgs pole inflation.

\subsection{Inflationary predictions}

We take the inflaton potential in eq.~(\ref{inflatonpot}) in the following,
\bea
V_E(\phi)=V_I \bigg[  \tanh\Big(\frac{\phi}{\sqrt{6}M_P}\Big)\bigg]^{2m},
\eea
with $V_I\equiv 3^m c_m\Lambda^{4-2m}M^{2m}_P$.
Then, we first obtain the slow-roll parameters,
\bea
\epsilon &=&\frac{1}{2} M^2_P \bigg(\frac{V'_E}{V_E}\bigg)^2 \nonumber \\
&=& \frac{4}{3} m^2 \bigg[  \sinh\Big(\frac{2\phi}{\sqrt{6}M_P}\Big)\bigg]^{-2} ,  \label{ep} \\
\eta &=& M^2_P \frac{V^{\prime\prime}_E}{V_E} \nonumber \\
&=&-\frac{4m}{3} \bigg[  \cosh\Big(\frac{2\phi}{\sqrt{6}M_P}\Big)-2m \bigg]  \bigg[\sinh\Big(\frac{2\phi}{\sqrt{6}M_P}\Big)\bigg]^{-2}. \label{eta}
\eea
The number of efoldings is 
\bea
N&=&\frac{1}{M_P} \int^{\phi_*}_{\phi_e} \frac{ {\rm sgn} (V'_E)d\phi}{\sqrt{2\epsilon}} \nonumber \\
&=&\frac{3}{4m}\, \bigg[ \cosh\Big(\frac{2\phi_*}{\sqrt{6}M_P}\Big)- \cosh\Big(\frac{2\phi_e}{\sqrt{6}M_P}\Big)  \bigg] \label{efold}
\eea
where $\phi_*, \phi_e$ are the values of the Higgs boson at horizon exit and the end of inflation, respectively. Here, we note that $\epsilon=1$ determines $\phi_e$.
As a result, using  eqs.~(\ref{ep}), (\ref{eta}) and (\ref{efold}) and $N\simeq \frac{3}{4m}\,  \cosh\Big(\frac{2\phi_*}{\sqrt{6}M_P}\Big)$ for $\phi_*\gg \sqrt{6} M_P$ during inflation, we obtain the slow-roll parameters at horizon exit in terms of the number of efoldings as
\bea
\epsilon_* &\simeq & \frac{3}{4\big(N^2-\frac{9}{16m^2}\big)},  \\
\eta_* &\simeq& \frac{3-2N}{2\big(N^2-\frac{9}{16m^2}\big)}.
\eea
Thus, we get the spectral index in terms of the number of efoldings, as follows, 
\bea
n_s&=&1-6\epsilon_*+2\eta_* \nonumber \\
&=& 1-\frac{4N+3}{2\big(N^2-\frac{9}{16m^2}\big)}. \label{sindex}
\eea
Moreover, the tensor-to-scalar ratio at horizon exit is
\bea
r=16\epsilon_* =  \frac{12}{N^2-\frac{9}{16m^2}}. \label{ratio}
\eea
Therefore, both the spectral index and the tensor-to-scalar ratio depend on the power of the Higgs potential, $m$, but the inflationary predictions are insensitive to the value of $m$.

As a result, from eq.~(\ref{sindex}), we obtain the spectral index as $n_s=0.966$ for $N=60$, which agrees with the observed spectral index from Planck, $n_s=0.967\pm 0.0037 $ \cite{planck}.
Moreover, we also predict  the tensor-to-scalar ratio as $r=0.0033$ for $N=60$, which is again compatible with  the bound from the combined Planck and Keck data \cite{keck}, $r<0.036$.

The CMB normalization, $A_s=\frac{1}{24\pi^2} \frac{V_I}{\epsilon_* M^4_P}=2.1\times 10^{-9}$,  sets the inflation energy scale by
\bea
3^m c_m \Big(\frac{\Lambda}{M_P}\Big)^{4-2m}= (3.1\times 10^{-8}) \,r=1.0\times 10^{-10}, \label{CMB}
\eea
where we took $r=0.0033$, in the second equality.
Thus, for $\Lambda\simeq M_P$, the coefficient $c_m$ of the effective Higgs interaction is constrained to be a small value for the CMB normalization. 

We note that for $m=0$, the inflaton potential in Einstein frame becomes constant, so it corresponds to a constant vacuum energy present by today. So, we assume that the constant vacuum energy is negligible for the graceful exit from inflation and it is tuned to make the effective vacuum energy small from the observation of the accelerating universe at present. 

For $m=1$ in eq.~(\ref{inflatonpot}),  the inflaton potential takes the same form as in Starobinsky inflation model. 
However, in this case, we can identify the coefficient of the leading Einstein-frame potential by $c_1 \Lambda^2=\mu^2_H$ in eq.~(\ref{HEFT}), which must be positive for inflation, unlike the tachyonic mass required for electroweak symmetry breaking.  Moreover, the vacuum energy during inflation is given by $V_I=3\mu^2_H M^2_P$, which is constrained by the CMB normalization. Thus, we need to take $\mu_H= 1.4\times10^{13}\,{\rm GeV}$, which is much larger than the Higgs parameter at low energy.  Therefore, for the SM Higgs inflation, we need to take the case with $m>1$.

For $m=2$ in eq.~(\ref{inflatonpot}), we identify the coefficient of  the leading Einstein-frame potential by the Higgs quartic coupling with $c_2=\lambda_H$ in eq.~(\ref{HEFT}) and the vacuum energy during inflation becomes $V_I=9 c_2M^4_P=9\lambda_H M^4_P$. In this case, the CMB normalization fixes $\lambda_H=1.1\times 10^{-11}$.

\subsection{General Higgs potential and inflationary bounds}

For electroweak symmetry breaking, we need to take the Higgs potential in Einstein frame to be a general form of the polynomial, as follows,
\bea
V_E(H) = V_0+ \mu^2_H |H|^2 +\lambda_H |H|^4 + \sum^\infty_{m=3} c_m \Lambda^{4-2m} |H|^{2m}, 
\eea
which corresponds to the one in the Jordan frame,
\bea
V_J(H) = \bigg(V_0+ \mu^2_H |H|^2 +\lambda_H |H|^4 + \sum^\infty_{m=3} c_m \Lambda^{4-2m} |H|^{2m} \bigg) \bigg(1-\frac{1}{3M^2_P}|H|^2\bigg)^2.
\eea
Then, we can rewrite the above potential in terms of the canonical Higgs field $\phi$ in unitary gauge, given in eq.~(\ref{can}), as follows,
\bea
V_E(\phi)&=& V_0 +3\mu^2_HM^2_P   \tanh^2\Big(\frac{\phi}{\sqrt{6}M_P}\Big)+9\lambda_HM^4_P  \tanh^4\Big(\frac{\phi}{\sqrt{6}M_P}\Big)\nonumber \\
&&+ \sum^\infty_{m=3} 3^m c_m \Lambda^{4-2m} M^{2m}_P   \tanh^{2m}\Big(\frac{\phi}{\sqrt{6}M_P}\Big).
\eea
As a result, there is a wider parameter space with $V_0\simeq 0$ for a successful inflation because the potential becomes almost constant near the Higgs pole.  Thus, setting $V_0=0$, we can generalize the CMB normalization in eq.~(\ref{CMB})  to
\bea
\frac{3\mu^2_H}{M^2_P} +9\lambda_H+ \sum^\infty_{m=3} 3^m c_m \Big(\frac{\Lambda}{M_P}\Big)^{4-2m}=1.0\times 10^{-10}. \label{gCMB}
\eea
Unless there are fine-tuned cancellations, we need to choose the individual terms in the Higgs potential to be small.

For the successful inflation with the Higgs quartic term, we need to take $9\lambda_H M^4_P\gtrsim 3|\mu^2_H|M^2_P, V_0$ during inflation. Then, we find it sufficient to take $\lambda_H\gtrsim  \frac{|\mu^2_H|}{3M^2_P}=1.1 \times 10^{-15}$ for the correct electroweak symmetry breaking \footnote{It is interesting to notice that the CMB normalization only leads to the bound on the Higgs mass parameter as $|\mu_H|\lesssim 1.4\times 10^{13}\,{\rm GeV}$, although we need to understand the Higgs mass hierarchy.} with $|\mu_H|=88\,{\rm GeV}$; $\lambda_H\gtrsim  \frac{V_0}{9M^2_4}=4.6 \times 10^{-85}$ for the observed dark energy, $V_0=(2.2\times 10^{-3}\,{\rm eV})^4$. Therefore, the CMB constraint on $\lambda_H$, namely,  $\lambda_H= 1.1\times 10^{-11}$, obtained in the previous subsection, is consistent with electroweak symmetry breaking and dark energy. Such a tiny Higgs quartic coupling during inflation is achievable due to the RG running of the Higgs quartic coupling, although it is subject to the precise values of the low energy parameters in the SM such as the top quark Yukawa coupling \cite{RGHiggs} and the threshold or running contributions from new particles \cite{threshold}.

We also remark that a successful inflation from the higher order Higgs potential with $m\geq 3$ is also possible, as far as $3^m c_m\big(\frac{\Lambda}{M_P}\big)^{4-2m}\gtrsim 9\lambda_H, \frac{3|\mu^2_H|}{M^2_P} , \frac{V_0}{M^4_P}$. In this case, from the CMB normalization in eq.~(\ref{CMB}), the Higgs quartic term is bounded by $\lambda_H\lesssim 1.1\times 10^{-11}$, whereas the observed values of the other renormalizable parameters, $|\mu^2_H|$ and $V_0$, are small enough. 
Furthermore, for the dominance with a single higher order term with $m$,  even higher order terms with $m+k$, $k\geq 1$, are suppressed for $c_m\gtrsim 3^k \big(\frac{M_P}{\Lambda}\big)^{2k} c_{m+k}$, but they are allowed for the slow-roll inflation, as far as the generalized CMB normalization in eq.~(\ref{gCMB}) is satisfied.

\section{Reheating}

We consider the inflaton condensate and the general equation of state during reheating.
Then, we first discuss the perturbative reheating from the decay or scattering of the Higgs inflaton into the SM particles and determine the minimal reheating temperature due to the production of the SM fermions. We also comment on the preheating and its effects on reheating through the production of heavy fermions and gauge bosons in the SM.

\subsection{Inflaton condensate}

First, for $|H|\ll \sqrt{6}M_P$, we can approximate the Higgs kinetic term in Einstein frame in a canonical form, that is, $\phi\simeq h$, so the Einstein-frame potential for the pole inflation in eq.~(\ref{inflatonpot}) takes
\bea
V_E(\phi)\simeq \frac{c_m}{2^m} \Lambda^{4-2m} \phi^{2m}\equiv \alpha_m \phi^{2m}, \label{rehpot}
\eea
which is anharmonic for $m>1$, leading to the general equation of state during reheating. 
After the period of exponential expansion, the inflaton $\phi$ begins to oscillate about the minimum of the potential. 
The end of inflation may be defined when $\ddot a=0$, where $a$ is the cosmological scale factor. The inflaton field value at that time is given by \cite{Ellisreheating, Garcia:2020wiy}
\begin{align}
   \phi_{\rm end} &\simeq 
   \sqrt{\frac{3}{8}}M_P\ln\left[\frac{1}{2} + \frac{2m}{3}\left(2m + \sqrt{4m^2+3}\right)\right].
   \label{phiend}
\end{align}
We can show that at the end of inflation, 
the condition $\ddot a=0$ is equivalent to $\dot\phi_{\rm end}^2 = V_E(\phi_{\rm end})$
and thus the inflaton energy density at $\phi_{\rm end}$ is  $\rho_{\rm end}=\frac{3}{2}V_E(\phi_{\rm end})$.

We now consider the averaged energy density and pressure for the inflaton and the equation of state during reheating. Applying the virial theorem to $S\equiv \phi {\dot\phi}$ by
\bea
\Big\langle\frac{dS}{dt} \Big\rangle=0,
\eea
we obtain
\bea
\Big\langle \frac{1}{2}{\dot\phi}^2\Big\rangle=-\Big\langle \frac{1}{2} \phi {\ddot\phi} \Big\rangle=m\langle V_E(\phi)\rangle.
\eea
Here, in the last equality, we used the equation of motion for the inflaton, ${\ddot\phi}=-V'$, by ignoring the Hubble friction term, and the inflaton potential during reheating from eq.~(\ref{rehpot}). As a result, the averaged energy density and pressure for the inflaton becomes
\bea
\rho_\phi &=&\Big\langle \frac{1}{2}{\dot\phi}^2\Big\rangle+\langle V_E(\phi)\rangle=(m+1) \langle V_E(\phi)\rangle, \label{infdens} \\
p_\phi &=& \Big\langle \frac{1}{2}{\dot\phi}^2\Big\rangle-\langle V_E(\phi)\rangle=(m-1) \langle V_E(\phi)\rangle.
\eea
Then, the averaged equation of state for the inflaton during reheating is given by
\bea
\langle w_\phi\rangle=\frac{p_\phi}{\rho_\phi} =\frac{m-1}{m+1}.  \label{eos1}
\eea
As a result, we get a general equation of state for the Higgs inflaton, which is different from the one for matter, $w_\phi\neq 0$, for $m\neq 1$. If the reheating process is not instantaneous, it is important for the reliable inflationary predictions to take into account the correct equation of state during reheating \cite{reheating}.

We take the inflaton to be $\phi=\phi_0(t) {\cal P}(t)$, where $\phi_0(t)$ is the amplitude of the inflaton oscillation and it is constant over oscillation, and  ${\cal P}(t)$ is the periodic function.  From eq.~(\ref{infdens}) with the energy conservation, we obtain $\rho_\phi=(m+1) \langle V_E(\phi)\rangle=V_E(\phi_0)$. Then, we get $\langle {\cal P}^{2m}\rangle=\frac{1}{m+1}$. 

From the energy density for the inflaton,
\bea
 \frac{1}{2}{\dot\phi}^2+V_E(\phi)=V_E(\phi_0),
\eea
we obtain the equation for the periodic function ${\cal P}$, as follows,
\bea
{\dot{\cal P}}^2 =\frac{2\rho_\phi}{\phi^2_0} \Big(1-{\cal P}^{2m}\Big)= \frac{m^2_\phi}{m(2m-1)}\,  \Big(1-{\cal P}^{2m}\Big) \label{Peq}
\eea
where we used the effective inflaton mass in the second equality,
\bea
m^2_\phi =V^{\prime\prime}_E(\phi_0) =2\alpha_m  m (2m-1) \phi^{2m-2}_0. \label{inflatonmass}
\eea
Thus, from the integral of  eq.~(\ref{Peq}), we get the angular frequency of the inflaton oscillation \cite{Garcia:2020wiy} as
\bea
\omega = m_\phi\sqrt{\frac{\pi m}{2m-1}}\, \frac{\Gamma\big(\frac{1}{2}+\frac{1}{2m}\big)}{\Gamma\big(\frac{1}{2m}\big)}. \label{freq}
\eea
As a result, we can make a Fourier expansion of the periodic function  $\cal P$ by
\bea
{\cal P}(t) =\sum_{n=-\infty}^\infty {\cal P}_n \,e^{-in\omega t}. 
\eea

\subsection{Boltzmann equations for reheating}

Including the effects of the Hubble friction and the inflaton decay, we find the equation of motion for the inflaton, as follows,
\bea
{\ddot\phi}+ (3H+\Gamma_\phi) {\dot\phi} +V'_E=0
\eea
where $\Gamma_\phi$  is the inflaton decay or scattering rate, given by
\bea
\Gamma_\phi=\sum_f\Gamma_{\phi\to f{\bar f}} + \sum_{V=W,Z}\Gamma_{\phi\phi\to VV}.
\eea
The above equation can be approximated to the Boltzmann equation for the averaged energy density,
\bea
{\dot\rho}_\phi + 3(1+w_\phi)H \rho_\phi\simeq -\Gamma_\phi (1+w_\phi) \rho_\phi.
\label{Boltzmann_phi}
\eea
Moreover, the Boltzmann equation governing the radiation energy density $\rho_R$ is given by
\bea
{\dot\rho}_R + 4 H\rho_R =\Gamma_\phi (1+w_\phi) \rho_\phi. \label{Boltzmann_rad}
\eea

First, due to the Yukawa couplings of the Higgs inflaton to the SM fermions, ${\cal L}_{\rm int}=-\frac{1}{\sqrt{2}}y_f\phi {\bar f} f$, for the inflaton condensate, $\phi=\phi_0(t) {\cal P}(t)$, we obtain the decay rate of the inflaton condensate \cite{Ichikawa:2008ne,Garcia:2020wiy}, as follows,
\bea
\Gamma_{\phi\to f{\bar f}}=\frac{1}{8\pi (1+w_\phi)\rho_\phi} \sum_{n=-\infty}^\infty |M^f_n|^2 (E_n\beta^f_n)
\eea
where $E_n=n \omega$ and
\bea
|M^f_n|^2&=&y^2_f \phi^2_0 |{\cal P}_n|^2 E^2_n \beta^2_n, \\
\beta^f_n &=& \sqrt{1-\frac{4m^2_f}{E^2_n}}. 
\eea
Then, averaging over oscillations, we get
\bea
\langle\Gamma_{\phi\to f{\bar f}}\rangle&=&\frac{y^2_f \phi^2_0 \omega^3}{8\pi (1+w_\phi)\rho_\phi}\, \sum_{n=-\infty}^\infty  n^3 |{\cal P}_n|^2 \langle\beta^3_n\rangle \nonumber \\
&=& \frac{y^2_f \omega^3}{8\pi m^2_\phi}\,(m+1)(2m-1) \Sigma_m^f \Bigg\langle\bigg(1-\frac{4m^2_f}{\omega^2 n^2}\bigg)^{3/2}\Bigg\rangle \label{decayrate}
\eea
where $\Sigma_m^f =\sum_{n=1}^\infty  n^3 |{\cal P}_n|^2$ whose values  are given as a function of the parameter $m$ in Table \ref{table:1}.
Here, the effective mass for the fermion is given by $m_f=\frac{1}{\sqrt{2}} y_f\phi(t)=(m_f/v)\phi(t)$ where $v$ is the VEV of the Higgs field in the true vacuum. 

Similarly, due to the gauge interactions of the Higgs inflaton to $W $ or $Z$ bosons, ${\cal L}_{\rm int}=\frac{1}{4}g^2 \phi^2 W^+_\mu W^{-\mu}+\frac{1}{8} (g^2+g^{\prime 2})\phi^2 Z_\mu Z^\mu$, in unitary gauge, we obtain the scattering rate of the inflaton condensate, as follows,
\bea
\Gamma_{\phi\phi\to VV}=\frac{1}{8\pi (1+w_\phi)\rho_\phi} \sum_{n=1}^\infty |M^V_n|^2 (E_n\beta^V_n)
\eea
with 
\bea
|M^W_n|^2&=&\frac{1}{4}g^4 \phi^4_0 |({\cal P}^2)_n|^2 \bigg(3+\frac{4E^4_n}{m^4_W}-\frac{4E^2_n}{m^2_W}\bigg), \\
|M^Z_n|^2&=&\frac{1}{8}(g^2+g^{\prime 2})^2 \phi^4_0 |({\cal P}^2)_n|^2 \bigg(3+\frac{4E^4_n}{m^4_Z}-\frac{4E^2_n}{m^2_Z}\bigg),
\eea
and $\beta^V_n = \sqrt{1-\frac{m^2_V}{E^2_n}}$ with $V=W, Z$. Here, the effective masses for the gauge bosons are given by $m^2_W=\frac{1}{4} g^2 \phi^2(t)$ and $m^2_Z=\frac{1}{4}(g^2+g^{\prime 2}) \phi^2(t)$, and $({\cal P}^2)_n$ are the Fourier coefficients of the expansion, ${\cal P}^2=\sum_{n=-\infty}^\infty ({\cal P}^2)_n \,e^{-in\omega t}$.
Then, the averaged scattering rate for the inflaton is given by
\bea
\langle\Gamma_{\phi\phi\to WW}\rangle&=&\frac{g^4 \phi^2_0 \omega}{16\pi m^2_\phi}\,(m+1)(2m-1) \Phi_W, \\
\langle\Gamma_{\phi\phi\to ZZ}\rangle&=&\frac{(g^2+g^{\prime 2})^2 \phi^2_0 \omega}{32\pi m^2_\phi}\,(m+1)(2m-1) \Phi_Z ,
\eea
with
\bea
\Phi_V\equiv \Sigma_m^V \bigg\langle\frac{\beta^V_n(3+3(\beta^V_n)^4-2(\beta^V_n)^2) }{(1-(\beta^V_n)^2)^2} \bigg\rangle, \quad V=W, Z
\eea
where $\Sigma_m^V=\sum_{n=1}^\infty n  |({\cal P}^2)_n|^2 $ whose values are given as a function of the parameter $m$ in Table \ref{table:1}.

\begin{table}[htbp!]
    \centering
    \begin{tabular}{|c|c|c|}
    \hline
        $m$  & $\Sigma_m^f$  & $\Sigma_m^V$ \\
        \hline
        1 &  0.250 & 0.125\\
        \hline
        2 &   0.241 &   0.125   \\
        \hline
        3 &    0.244    &    0.124  \\
        \hline
        4 &    0.250       &     0.122 \\
        \hline
        5 &     0.257      &       0.120        \\
        \hline
        6 &  0.264        &    0.119 \\
        \hline 
        7 &    0.270      &    0.117      \\
        \hline
        8  &     0.276      &     0.116      \\
        \hline
        9 &      0.281        &       0.115      \\
        \hline
        10 &      0.287        &       0.114       \\
        \hline
        
    \end{tabular}
    \caption{Sums of Fourier coefficients, $\Sigma_m^{f,V}$, appearing in the decay rates of the Higgs inflaton. We chose some values of the equation of state parameter $m$ during reheating.}
    \label{table:1}
\end{table}

We note that the decay and scattering rates of the inflaton scale with the inflaton energy density by $\Gamma_{\phi\to f{\bar f}}=\gamma_\phi \rho^l_\phi$ with $l=\frac{1}{2}-\frac{1}{2m}$ and $\Gamma_{\phi\phi\to VV}={\hat\gamma}_\phi \rho^n_\phi$ with $n=\frac{3}{2m}-\frac{1}{2}$  \cite{Garcia:2020wiy}. For instance, for the quadratic potential with $m=1$, we get $l=0$ and $n=1$, so the scattering rate is smaller than the decay rate for similar inflaton couplings. In this case, the
scattering process cannot achieve the reheating mechanism alone, without any decay of the
inflation background. However, for $m=2$, both the decay and scattering rates scale by $\rho^\frac{1}{2}_\phi$, so they are comparable. Moreover, the $m>2$ case makes the decay rate being of higher power in the inflaton energy density, that is, $l>n$, tending to be suppressed as compared to the scattering rate. 

 From eq.~(\ref{freq}) with eqs.~(\ref{inflatonmass}) and eq.~(\ref{rehpot}), we have $\omega^2\sim m^2_\phi\sim c_m \Lambda^{4-2m}\phi^{2m-2}_0$. Then, together with $m_f\sim y_f \phi_0$ and $m_V\sim g \phi_0$, we get $m^2_f/\omega^2\sim (y^2_f/c_m)(\phi_0/\Lambda)^{4-2m}\lesssim 1$ and $m^2_V/\omega^2\sim (g^2/c_m)(\phi_0/\Lambda)^{4-2m}\lesssim 1$. Then, for $c_m\sim 10^{-10}$ from the CMB normalization in eq.~(\ref{CMB}),  we can avoid the kinematic suppression for the inflaton decay or scattering, only if $y_f, g\lesssim \sqrt{c_m}\sim 10^{-5}(\Lambda/\phi_0)^{4-2m}$. 
 
 The effect of the time-dependent effective masses for fermions and bosons can be determined by averaging over the oscillations the effective decay or scattering rates. The phase-space dependence on the effective masses leads to a suppression of the production rates, even if the production is never completely blocked, as for any coupling there will be a time interval around $\phi=0$ during which the production is allowed \cite{Garcia:2020wiy}. The resulting suppression in the production for large couplings to the inflaton can be numerically evaluated. 
 
 In the initial stage of the inflaton oscillation with $\phi_0\sim \Lambda\sim M_P$ just after inflation, as in eq.~(\ref{phiend}),  only light quarks and leptons in the SM can be produced abundantly from the perturbative decays of the Higgs inflaton. However, heavy quarks or gauge bosons can be produced only in the small field values with $\phi_0\ll \Lambda$ (or at the later time) for $m<2$ and in the large field values with $\phi_0\sim\Lambda$ (or at the earlier time) for $m>2$, when the decay or scattering channels of the inflaton are kinematically open. Thus, as will be discussed in the later section, the parametric resonance during preheating could be also relevant for the gauge boson production.

\subsection{Reheating temperature}

For $a_{\rm end} \ll a\ll a_{\rm RH}$ where $a_{\rm RH}$ is the scale factor at the time reheating is complete, we can ignore the inflaton decay rate and integrate  eq.~(\ref{Boltzmann_phi}) without the decay rate to obtain
\be
\rho_{\rm \phi}(a)\simeq \rho_{\rm end}\left(\frac{a_{\rm end}}{a}\right)^{\frac{6m}{m+1}}. \label{inflatondensity}
\ee
This is due to the general equation of state during reheating, given in eq.~(\ref{eos1}).
When the reheating process is dominated by the perturbative decays of the inflaton into light fermions in the SM, we obtain the reheating  temperature, depending on the power of the leading inflaton potential, $V(\phi)=\alpha_m \phi^{2m}$, during reheating, as follows  \cite{Garcia:2020wiy},
\bea
T_{\rm RH}= \left\{\begin{array}{cc} \bigg(\frac{30}{\pi g_*(T_{\rm RH})}\bigg)^{1/4} \bigg[ \frac{2m}{7-2m}\, \sqrt{3}M_P \gamma_\phi\bigg]^{\frac{m}{2}}, \qquad\qquad 7-2m>0, \vspace{0.3cm}\\  \bigg(\frac{30}{\pi g_*(T_{\rm RH})}\bigg)^{1/4}\bigg[\frac{2m}{2m-7}\,\sqrt{3}M_P\gamma_\phi\,( \rho_{\rm end})^{\frac{2m-7}{6m}}\bigg]^{\frac{3m}{4(m-2)}}, \quad 7-2m<0 \end{array} \right.
\eea
where we approximated the decay rate of the inflaton from eq.~(\ref{decayrate}) by $\Gamma_\phi\simeq \Gamma_{\phi\to f{\bar f}} \equiv \gamma_\phi \rho^l_\phi$, with $l=\frac{1}{2}-\frac{1}{2m}$, and
\bea
\gamma_\phi\equiv  \sum_f \frac{1}{8} N_c y^2_f\sqrt{2\pi}  m^2 (m+1)\, (\alpha_m)^{\frac{1}{2m}}\,\bigg(\frac{\Gamma\big(\frac{1}{2}+\frac{1}{2m}\big)}{\Gamma\big(\frac{1}{2m}\big)}\bigg)^3 \Sigma^f_m
\eea
where $f$ runs over all the SM fermions that can be produced and $N_c$ indicates the number of colors for each fermion.
Here, $g_*(T_{\rm RH})$ is the effective number of degrees of freedom at the reheating temperature $T_{\rm RH}$, and $\rho_{\rm end}$ is the inflaton energy density at the end of inflation, and we have neglected the average effective masses for the light fermions generated by their couplings to the inflaton during reheating.
We note that for $2m<7$, the reheating temperature  is independent of $\rho_{\rm end}$, whereas for $2m>7$, the reheating temperature depends on $\rho_{\rm end}$.

We now consider the effects of  time-dependent effective masses of fermions during the decay process of the Higgs inflaton. These effects are included in the calculations of the decay rate (\ref{decayrate}) with the averaged effective kinetic factors. For small effective masses, we can neglect the kinematic factors. However, for large Yukawa couplings to the Higgs inflaton, the production of the fermion is suppressed, but not completely blocked. In the limit of large effective masses, we can replace the averaged kinetic factor in the decay rate by 
\bea
\Gamma_\phi &=& \gamma_\phi\rho_\phi^l\Bigg\langle\bigg(1-\frac{4m^2_f}{\omega^2 n^2}\bigg)^{3/2}\Bigg\rangle  \nonumber \\
& \simeq&  \gamma_\phi\rho_\phi^l\times\mathcal{R}^{-1/2},
\eea
with $\mathcal{R} = \left(\frac{2m_f}{\omega}\right)^2_{\phi\rightarrow \phi_0}$,  where  the approximation in the second line is valid if $\mathcal{R}\gg1$ \cite{Garcia:2020wiy}. We consider only the first Fourier mode of the Higgs inflaton background in the effective masses of fermions as the dominant contribution. The higher order modes also contribute with much smaller Fourier coefficients, making their contributions negligible up to $\sim 10\%$ level. We performed the numerical integration of the Boltzmann equations in eqs.~(\ref{Boltzmann_phi}) and (\ref{Boltzmann_rad}), taking into account the averaged kinetic factor in the limit $\mathcal{R}\gg 1$ for heavy fermions during the integration.  As a result, we determine the reheating temperature from the perturbative decays of the Higgs inflaton into fermions, as given in Table \ref{table:2}. 

\begin{table}[htbp!]
    \centering
    \begin{tabular}{|c|c|}
    \hline
       $m$   &   $T_{\rm RH}$ $[\rm GeV]$    \\
    \hline
       1  &  $5.1\times10^{13}$  \\
    \hline
      2 &        $2.6\times10^{9}$ \\
      \hline
      3 &  260 \\
      \hline
      4 & $9.4\times 10^5$ \\
      \hline
      5 & $2.1\times 10^7$ \\
      \hline
      6 & $1.1\times 10^8$ \\
      \hline
      7 & $2.8\times10^8 $ \\
      \hline
      8 & $4.9\times10^8$ \\
    \hline   
    9 & $8.4\times10^8$ \\
    \hline
    10 & $1.2\times10^9$ 
    \\
    \hline
    \end{tabular}
    \caption{Reheating temperature $T_{\rm RH}$, determined from the decays of the Higgs inflaton into the SM fermions, including the kinematic suppression for the effective fermion masses. We chose some values of the equation of state parameter $m$ during reheating. }
    \label{table:2}
\end{table}

\subsection{Preheating}

We consider a non-perturbative process for reheating, the so called preheating, which leads to the non-adiabatic excitation of the perturbations for fields coupled to the Higgs inflaton. 

In the limit of sub-Planckian Higgs fields below the pole, we can approximate the Higgs perturbation, $\varphi_k$, with the comoving momentum $k$, as the following modified Klein-Gordon equation, 
\bea
{\ddot\varphi}_k + 3H {\dot \varphi}_k + \bigg(\frac{k^2}{a^2}+m^2_\varphi(t) +6\xi_H \Big(\frac{\ddot{a}}{a}+\frac{{\dot a}^2}{a^2}\Big)\bigg) \varphi_k =0 \label{higgspert}
\eea
where $m_\varphi(t)$ is  the effective mass  for the Higgs perturbation during reheating, $m^2_\varphi(t)=m^2_\phi(t) {\cal P}^{2m-2}$, with $m^2_\phi(t)$ being given by eq.~(\ref{inflatonmass}). Here, $\xi_H$ is the non-minimal coupling to the SM Higgs, which is chosen to $\xi_H=-\frac{1}{6}$ in our model. Then, we make a change of variables with $z=\omega t$ where $\omega$ is the angular frequency for the inflaton oscillation, given in eq.~(\ref{freq}).
Then, from eq.~(\ref{higgspert}), we get the equation for the rescaled perturbation, $H_k=\omega^{1/(1-m)} \varphi_k$, as
\bea
H^{\prime\prime}_k+   \bigg(\kappa^2+\frac{ m^2 m^2_\varphi(t)}{\omega^2}+\frac{2\xi_H(m+1)(2-m)}{3(1-m)^2} \cdot\bigg(\frac{\omega'}{\omega}\bigg)^2 \bigg) H_k=0 \label{pert2}
\eea
where 
\bea
\kappa^2\equiv \frac{m^2 k^2}{\omega^2 a^2},
\eea
the prime denotes the derivative with respect to $z$ with $dz=\frac{\omega}{m}\, dt$, and we used ${\dot\omega}=\frac{1-m}{m}\, \frac{\omega}{t}$. This is the master equation governing the generalized preheating with an anharmonic inflaton potential. We note that the last term proportional to $(\omega'/\omega)^2$ in eq.~(\ref{pert2}) is due to the Hubble expansion, so it can be ignored if the particles are produced efficiently for several oscillations of the inflaton condensate. 

Here, we remark that we have used the following information for deriving eq.~(\ref{pert2}). First, from the Friedmann equation with eq.~(\ref{inflatondensity}), the scale factor scales by $a\propto t^{\frac{m+1}{3m}}$ during reheating, so the Hubble parameter becomes $H=\frac{m+1}{3m}\, \frac{1}{t}$. Then, from the averaged Friedmann equation, $\langle H^2 \rangle=\frac{1}{3M^2_P}\, \langle \rho\rangle$, and the averaged inflaton density in eq.~(\ref{infdens}), $\langle \rho\rangle\propto \langle V_E(\phi)\rangle\sim \phi^{2m}_0$, the inflaton condensate scales by $\phi_0(t)\propto t^{-1/m}$, leading to $\omega\propto m_\phi\propto \phi^{m-1}_0\propto t^{1/m-1}$ and ${\dot\omega}=\frac{1-m}{m}\, \frac{\omega}{t}$. 

As a result, we find that the effective mass term in eq.~(\ref{pert2}) becomes
\bea
\frac{m^2m^2_\varphi(t)}{\omega^2}= \frac{1}{\pi}\, m(2m-1)\bigg(\frac{\Gamma\big(\frac{1}{2m}\big)}{\Gamma\big(\frac{1}{2}+\frac{1}{2m}\big)}\bigg)^2 {\cal P}^{2m-2}(t),
\eea
which does not depend on the  amplitude of the inflaton oscillation, $\phi_0$, but instead it depends on the power of the inflaton potential during reheating.  This case is similar to the scale-invariant model for inflation with a quartic inflaton model during reheating \cite{preheating,inflatondm}, namely, the $m=2$ case, but the oscillating part is different if $m\neq 2$.

Similarly, following the similar procedure as above, the perturbation equations for $W, Z$ gauge bosons are obtained from the one for the Higgs perturbation in eq.~(\ref{pert2}), after the effective mass $m^2_\varphi(t)$ is replaced by those for the gauge bosons, $m^2_V(t)$, with $V=W, Z$, and $\xi_H=0$. In this case, the effective mass term for the $W$ boson perturbation becomes
\bea
\frac{m^2m^2_W(t)}{\omega^2} = \frac{g^2}{8\pi \alpha_m}\, \bigg(\frac{\Gamma\big(\frac{1}{2m}\big)}{\Gamma\big(\frac{1}{2}+\frac{1}{2m}\big)}\bigg)^2 \phi^{4-2m}_0(t) {\cal P}^2(t),  \label{gaugemass}
\eea
which also depends on the  amplitude of the inflaton oscillation, $\phi_0$, unlike the case for the Higgs perturbation. 
For the effective mass term for the $Z$ boson perturbation, we only have to replace $g^2$ with $g^2+g^{\prime 2}$ in the above result.  

As a result, as $z=\omega t \propto t^{1/m}$, we get $\phi_0\propto t^{-1/m}\propto z^{-1}$, so the effective mass term for the gauge boson perturbation depends on the amplitude of the inflaton oscillation because $ \phi^{4-2m}_0(t) \propto z^{2m-4}$.  Therefore, for $m<2$, there can be a kinematic blocking only in the early time for the perturbative decay of the inflaton into heavy fermions or gauge bosons in the SM, so the preheating can be important in this case.  On the other hand, for $m>2$, the perturbative decay of the inflaton becomes important in the early time, but if it is forbidden, preheating can be again important for the production of heavy particles.  We also note that the coefficients in the effective mass terms for the gauge boson perturbations in eq.~(\ref{gaugemass}) are proportional to $g^2/\alpha_m$, which is large for $\alpha_m\lesssim 10^{-10}$ from the CMB normalization. Therefore, preheating could lead to a higher reheating temperature than the one obtained from the perturbative reheating, as in the original Higgs inflation with a large non-minimal coupling \cite{Higgsreheating}.

For the early period of the inflaton oscillation, the case with $m>2$ effectively introduces a large coupling to the inflaton as time goes by, so the parametric resonances for the particle production are extended to large momentum modes \cite{preheating,inflatondm}.  On the other hand,  for the case with $m<2$, the resonant production of particles can be limited to relatively smaller momentum modes. The detailed discussion on preheating and particle production is beyond the scope of our work, so it will be postponed to a forthcoming work \cite{progress}.

\section{Extensions of Higgs pole inflation}

In this section, we consider the general effective interactions for the Higgs fields in the general Higgs pole expansions and show the examples for a vanishing small Higgs quartic coupling during inflation from the running Higgs quartic coupling in the SM and its extension with a singlet scalar field. We also discuss embedding the Higgs pole inflation into supergravity where a general polynomial in the superpotential gives rise to the F-term potential for the Higgs pole inflation along the D-flat direction.

\subsection{Generalized Higgs pole expansions}

The condition for the pole inflation corresponds to a particular choice of the effective Jordan-frame potential in eq.~(\ref{HEFT}), with the only nonzero coefficients being given by
\bea
c_{m+1}=-\frac{2}{3} c_m \frac{\Lambda^2}{M^2_P}, \quad c_{m+2}=\frac{1}{9} c_m \frac{\Lambda^4}{M^4_P}.
\eea
However, we can regard the above choice of the Higgs potential as the leading order expansion near the Higgs pole, so our inflationary predictions are insensitive to the general Higgs pole expansions as will be discussed below. 

First, we generalize the non-minimal coupling function and the Higgs potential for the pole inflation, in the following,
\bea
\Omega(H)&=&\bigg(1-\frac{1}{3M^2_P}|H|^2\bigg)\sum_{n=0}^\infty{\tilde b}_n \bigg(1-\frac{1}{3M^2_P}|H|^2\bigg)^n,  \label{fo}\\
V_J(H)&=&\Lambda^{4-2m} |H|^{2m} \bigg(1-\frac{1}{3M^2_P}|H|^2\bigg)^2\sum_{n=0}^\infty {\tilde c}_n\bigg(1-\frac{1}{3M^2_P}|H|^2\bigg)^n \label{fv}
\eea
where we can take ${\tilde b}_0=1$ and ${\tilde c}_0=c_m$ without loss of generality. 
In this case, we also obtain the Einstein-frame potential as
\bea
V_E&=& \frac{V_J}{\Omega^2} \nonumber \\
&=&c_m \Lambda^{4-2m} |H|^{2m} \sum_{n=0}^\infty {\hat c}_n\bigg(1-\frac{1}{3M^2_P}|H|^2\bigg)^n \label{infpotg}
\eea
where ${\hat c}_n$'s are the redefined  coefficients in the pole expansion.
Then, near the Higgs pole, $|H|\sim\sqrt{3} M_P$, during inflation, we can ignore the higher order terms, $(1-\frac{1}{3M^2_P}|H|^2)^n$, with $n\geq 1$, in both the non-minimal coupling in eq.~(\ref{fo}) and the effective Higgs potential in eq.~(\ref{fv}). 
We note that there was a related discussion on the expansion of the Jordan frame potential for the inflaton in terms of the non-minimal coupling function, $\Omega=1-|H|^2/(3M^2_P)$, in Ref.~\cite{expansions}, which assumes the presence of degenerate vacua other than the origin of the inflaton.

Using the canonical Higgs field during inflation in eq.~(\ref{can}) near the Higgs pole, we can approximate the inflaton potential in eq.~(\ref{infpotg}) to
\bea
V_E(\phi)\simeq  3^m \Lambda^{4-2m} M^{2m}_P \bigg[  \tanh\Big(\frac{\phi}{\sqrt{6}M_P}\Big)\bigg]^{2m}\sum_{n=0}^\infty   {\hat c}_n\bigg[  \cosh\Big(\frac{\phi}{\sqrt{6}M_P}\Big)\bigg]^{-2n}. \label{gpole}
\eea

During the slow-roll inflation with $\phi\gg \sqrt{6} M_P$, the generalized inflaton potential in the pole expansion in eq.~(\ref{gpole}) becomes
\bea
V_E(\phi)&\simeq&  3^m \Lambda^{4-2m} M^{2m}_P \bigg(1-4m\,e^{-2\phi/(\sqrt{6}M_P)}\bigg) \sum_{n=0}^\infty 2^{-2n} {\hat c}_n e^{-2n\phi/(\sqrt{6}M_P)}\Big(1+e^{-2\phi/(\sqrt{6}M_P)}\Big)^{-2n} \nonumber \\
&\simeq & 3^m c_m  \Lambda^{4-2m} M^{2m}_P \bigg(1-\Big(4m\,-\frac{{\hat c}_1}{4c_m}\Big)\,e^{-2\phi/(\sqrt{6}M_P)}+\cdots\bigg). 
\eea
Then, as far as ${\hat c}_1\lesssim c_m$, the next order terms in the pole expansion can be subdominant for the inflaton potential, so our previous discussion based on the leading order term in the pole expansion is valid.

 \subsection{Running Higgs quartic coupling and Higgs pole inflation}

\begin{figure}[!t]
\begin{center}
\includegraphics[width=0.42\textwidth,clip]{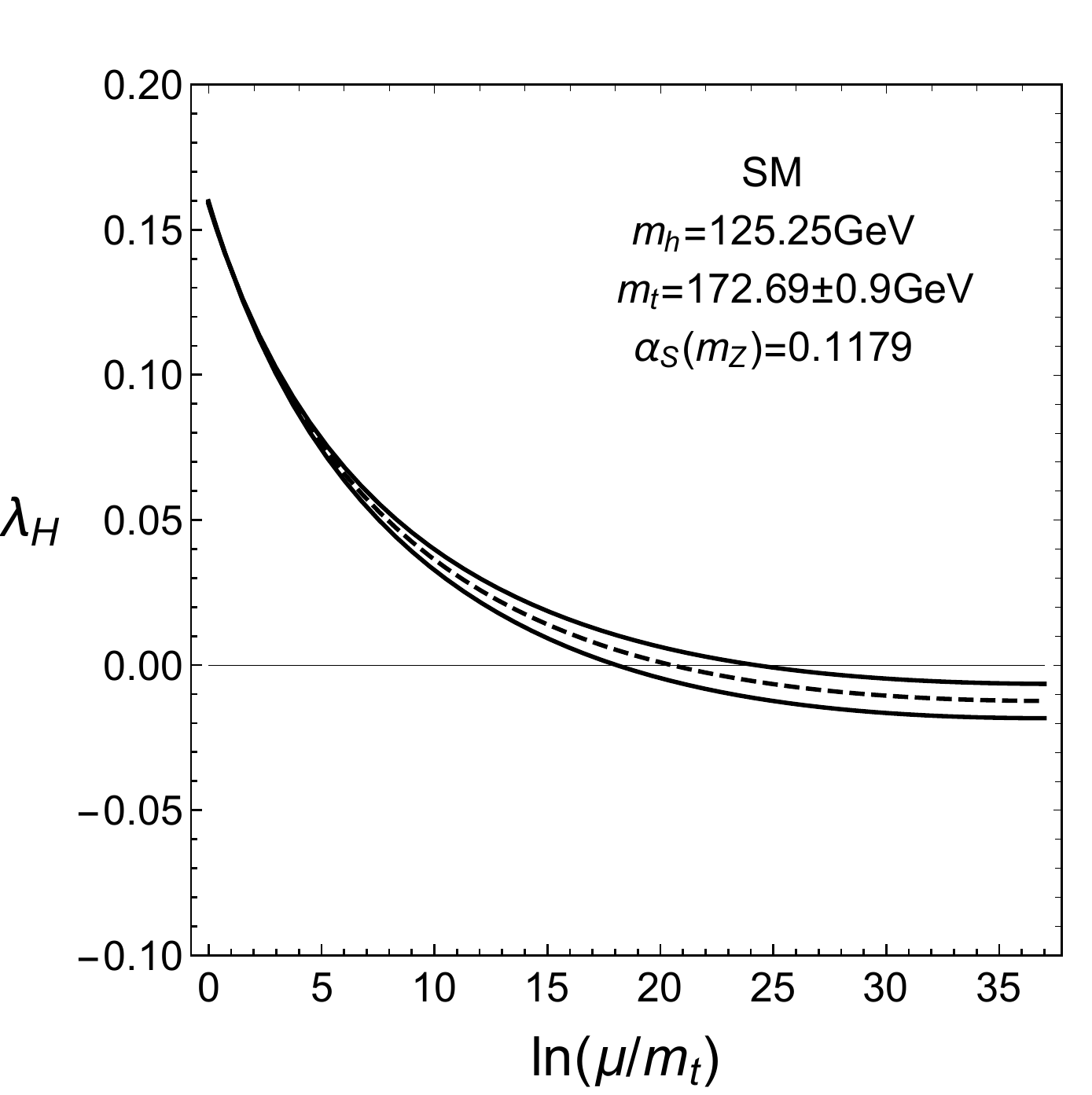}\,\, \includegraphics[width=0.42\textwidth,clip]{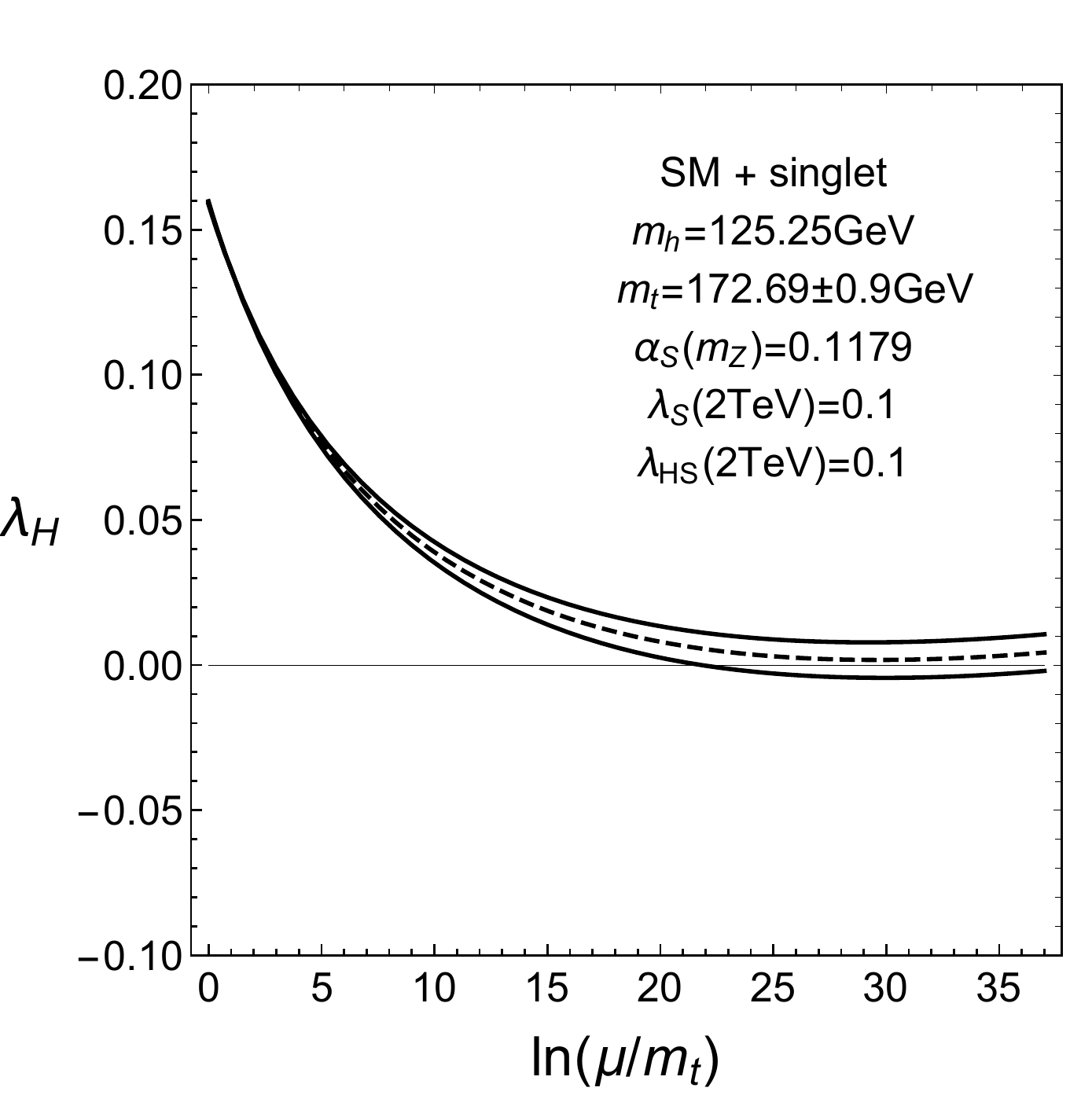} \,\, \includegraphics[width=0.42\textwidth,clip]{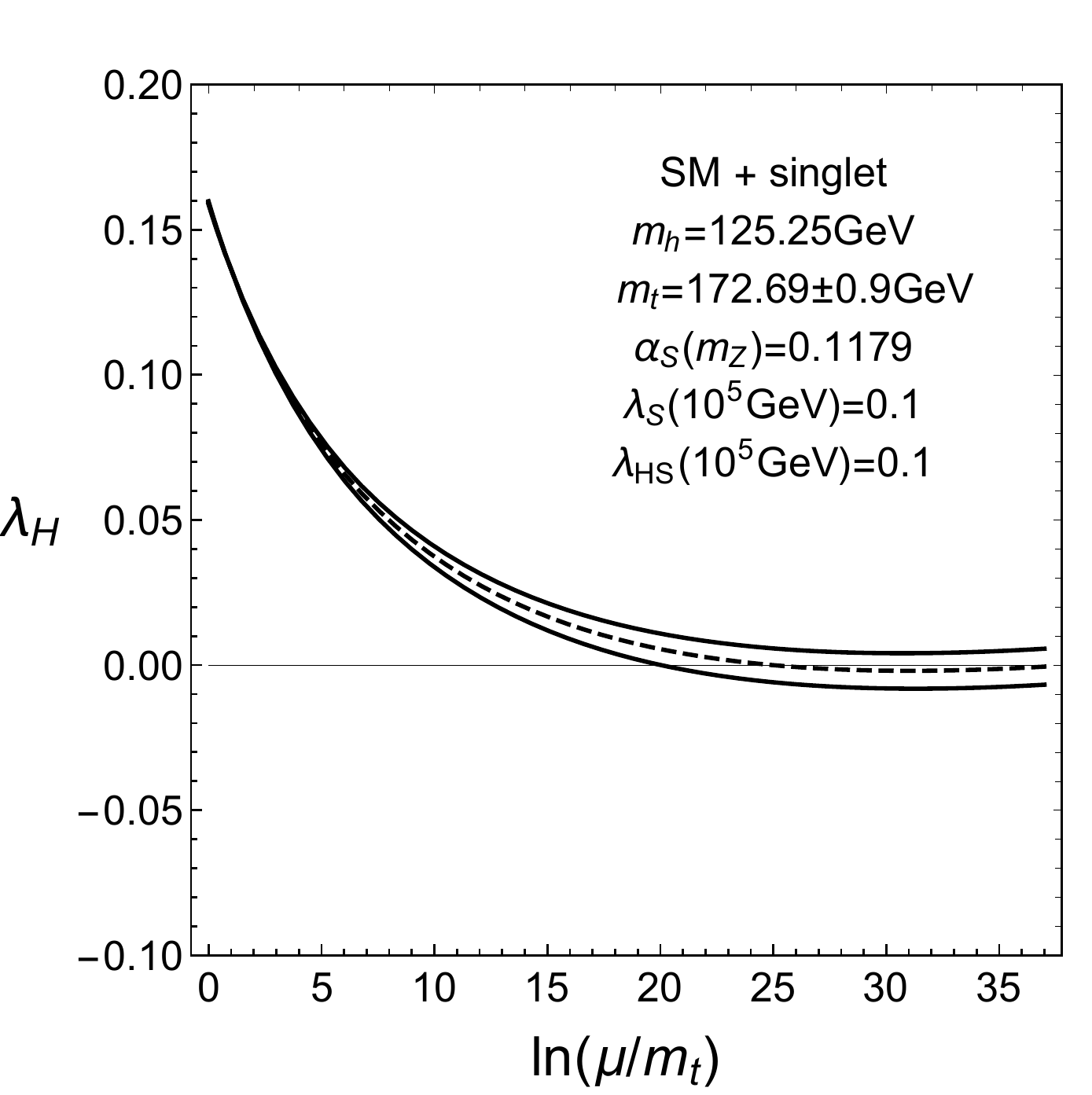}
\end{center}
\caption{Running Higgs quartic coupling as a function of the renormalization scale in units of $\ln(\mu/m_t)$ in the SM in the left plot in the upper panel, in the SM plus a complex singlet scalar field with $2\,{\rm TeV}$ and $10^5\,{\rm GeV}$ masses in the right plot in the upper panel and the lower panel, respectively. We took $m_h=125.25\,{\rm GeV}$, $m_t=(172.69\pm 0.90)\,{\rm GeV}$, $\alpha_S=0.1179$ for all the plots and $\lambda_S=\lambda_{HS}=0.1$ in the right plot in the upper panel and the lower panel.}
\label{fig:running}
\end{figure}

We illustrate the necessary positive small value for the Higgs quartic coupling during inflation in the extension of the SM with a complex singlet scalar field $S$. In this case, the one-loop renormalization group equations above the scalar mass $m_S$ are given \cite{threshold} by
\bea
(4\pi)^2 \frac{d\lambda_H}{d\ln\mu} &=& \Big(12y^2_t -3 g^{\prime 2}-9 g^2 \Big)\lambda_H -6y^2_t +\frac{3}{8} \bigg[2g^4+(g^{\prime 2}+g^2)^2 \bigg] +24\lambda^2_H +4 \lambda^2_{HS},  \nonumber \\
(4\pi)^2 \frac{d\lambda_{HS}}{d\ln\mu} &=& \frac{1}{2} \Big(12y^2_t-3g^{\prime 2}-9g^2 \Big)\lambda_{HS} +4\lambda_{HS}(3\lambda_H+2\lambda_S) + 8\lambda^2_{HS}, \\
(4\pi)^2 \frac{d\lambda_S}{d\ln\mu} &=& 20\lambda^2_S +8\lambda^2_{HS} \nonumber 
\eea
where the extra  quartic couplings are included in the Einstein frame potential in addition to the Higgs potential, 
\bea
V_E(S,H) = m^2_S |S|^2+ \lambda_S |S|^4 +2\lambda_{HS} |H|^2 |S|^2. 
\eea
The extra singlet scalar field becomes decoupled during inflation, because its effective mass $m^2_{S,{\rm eff}}=2\lambda_{HS}\langle|H|^2\rangle\sim \lambda_{HS} M^2_P$  is large enough for a sizable mixing quartic coupling, and it is held fixed to the origin for $\lambda_{HS}>0$. However, the mixing quartic coupling $\lambda_{HS}$ gives rise to a positive contribution to the beta function for the Higgs quartic coupling and can help maintain the Higgs quartic coupling all the way to the inflation scale, namely, $\mu\sim H\sim M_P$.

The running Higgs quartic coupling is sensitive to the Higgs mass, the top quark mass and the strong coupling, whose experimental values from Particle Data Group are $m_h=(125.25\pm 0.17)\,{\rm GeV}$, $m_t=(172.69\pm 0.30)\,{\rm GeV}$ and $\alpha_S=0.1179\pm 0.0009$ \cite{PDG}.
In the left plot in the upper panel of Fig.~\ref{fig:running}, we depict the running  Higgs quartic coupling as a function of the renormalization scale, $\ln(\mu/m_t)$, in the SM. We chose the Higgs mass to $m_h=125.25\,{\rm GeV}$, the top quark mass to $m_t=(172.69\pm 0.90)\,{\rm GeV}$ within the $3\sigma$ range, and $\alpha_S=0.1179$. In this case, the running Higgs quartic coupling turns negative around $\mu\sim 10^{10}-10^{11}\,{\rm GeV}$, depending on the top quark mass \cite{RGHiggs}.

On the other hand, in the right in the upper panel and the lower panel of Fig.~\ref{fig:running}, we also show the running Higgs quartic coupling as a function of the renormalization scale, $\ln(\mu/m_t)$, in the extension of the SM with a complex singlet scalar field. We chose the same set of the SM parameters as in the left plot of Fig.~\ref{fig:running} and took the new quartic couplings of the singlet scalar field to $\lambda_S=\lambda_{HS}=0.1$ at the singlet threshold $m_S=2\,{\rm TeV}, 10^5\,{\rm GeV}$, respectively. In this case, the running Higgs quartic coupling becomes positive due to the Higgs mixing quartic coupling all the way to the Planck scale \cite{threshold}, being consistent with the experimental constraints on the singlet scalar field beyond the TeV scale.  In this case, the resultant Higgs quartic coupling can be positive and small enough for the successful Higgs pole inflation.

We also remark that there is a natural possibility of choosing almost vanishing quartic couplings in the scalar potential during inflation near conformality. For instance, in the gauged $B-L$ or similar extensions of the Standard Model with right-handed neutrinos where the $B-L$ breaking scale as well as the electroweak scale are generated dynamically by the Coleman-Weinberg mechanism \cite{CW}. 

Finally, we make a comment on the possibility of a sufficiently small Higgs quartic coupling during inflation. First, under the Weyl scaling of the metric, $g_{E,\mu\nu}=\Omega g_{J,\mu\nu}$,  we need to rescale the SM fermions by $f'=\Omega^{-3/4} f$ for the canonical fermion kinetic terms and get the Yukawa couplings in the Einstein frame, as follows,
\bea
{\cal L}_Y&=&-\sqrt{-g_J}\, \frac{1}{\sqrt{2}} y_f h {\bar f} f \nonumber \\
&=&-\sqrt{-g_E} \, \frac{1}{\sqrt{2}} y_f \Omega^{-1/2} h {\bar f}' f'. 
\eea
Thus, the effective SM fermion masses in the Einstein frame become
 \bea
 M_f = \frac{1}{\sqrt{2}} y_f \varphi, \qquad  \varphi \equiv \frac{h}{\big(1-\frac{h^2}{6}\big)^{1/2}}.
 \eea
 Similarly, the effective masses for the $W$ and $Z$ bosons during inflation are given by
 \bea
 M_W =  \frac{1}{2} g \varphi, \qquad M_Z = \frac{1}{2}(g^2+g^2_Y)^{1/2}\varphi.
 \eea
 Then, both top quark and gauge boson masses depend on the redefined Higgs field $\varphi$, so they become close to Planckian values for $h\to \sqrt{6}$ or $\varphi\simeq \frac{\sqrt{6}}{2}\,e^{\phi/\sqrt{6}}\simeq 2\sqrt{m N}$,  during the Higgs pole inflation.
 As a result, the effective Higgs quartic term in the Einstein frame becomes
 \bea
 V_{E,{\rm eff}}(h) =\frac{1}{4} \lambda_H(\varphi(h)) h^4
 \eea
 where $\lambda_H(\varphi(h))$ is the running quartic coupling in the SM with the renormalization scale set to $\mu=\varphi$ in order to minimize the logarithms in the Coleman-Weinberg potential  \cite{RGHiggs,sigma,critical2}. 
 Then, as discussed in the previous section, for the successful Higgs pole inflation, we need to keep the running Higgs quartic coupling to satisfy $\lambda_H\lesssim 1.1\times 10^{-11}$ from the CMB normalization. Suppose that there is a minimum of the Higgs quartic coupling at the inflation scale. Then, the Higgs quartic coupling is expanded around the minimum at $\varphi=\varphi_*$, namely, around the scale with a vanishing beta function, as follows,
 \bea
 \lambda_H(\varphi)= \lambda_* +b \bigg(\ln \frac{\varphi}{\varphi_*}\bigg)^2,
 \eea 
 with $\lambda_*$ being the Higgs quartic coupling at $\varphi=\varphi_*$ and $b$ being the beta function coefficient of the Higgs quartic coupling at two-loops, given by $b=10^{-5}$ in the SM \cite{RGHiggs,reheating,critical2}.
 Then, as the Higgs field varies from $\varphi=\varphi_*$ to $\varphi=\varphi_e\sim 1$ during inflation, the running Higgs quartic coupling would get corrected significantly in the SM, being much larger than the CMB bound, so it would not be compatible with the Higgs pole inflation.
 However, the running Higgs quartic coupling would receive extra corrections easily in the extension of the SM with new scalar and gauge sectors. In this case, it is conceivable that even the two-loop and higher-loop beta function coefficients of the Higgs quartic coupling can be sufficiently small during inflation \cite{twoloops}, so a small Higgs quartic coupling could be maintained as required for the Higgs pole inflation. This possibility would correspond to a near-conformal theory replacing the SM in the UV, keeping the Higgs quartic coupling constant during inflation.  
 
 Another possibility to maintain a small Higgs quartic coupling during inflation is to introduce a dynamical mechanism for relaxing the effective Higgs quartic coupling to a small value. As will be discussed in the supersymmetric Higgs pole inflation in the next section, the extended Higgs sector with a certain symmetry could relax the effective Higgs quartic coupling to zero dynamically. In this case, there appears an extra Higgs potential of different origin, such as the F-term in the supersymmetric case. In particular, the higher order polynomial with a small coefficient can be used for the Higgs pole inflation.  
 
 We also remark that the Higgs quartic coupling is small at the onset of reheating as a result of the near-conformal theory or the dynamical relaxation mechanism in the extended Higgs sector in the UV, so the analysis of the perturbative reheating dynamics in the previous section is applicable. The concrete discussion on a particular realization of a small Higgs quartic coupling during inflation is beyond the scope of our work, so we leave this important issue as a future work.

\subsection{Higgs pole inflation in supergravity} 
 
We consider embedding the Higgs pole inflation into supergravity. For that, we introduce a modulus chiral multiplet $T$ and a pair of Higgs chiral multiplets, $H_u$ and $H_d$, which are conformally coupled to gravity in the Jordan frame. 
Then, we take the K\"ahler potential and the superpotential in the following form,
\bea
K&=&-3M^2_P\ln \Big(T+{\bar T}-\frac{1}{3M^2_P} |H_u|^2-\frac{1}{3M^2_P}|H_d|^2 \Big)\equiv -3M^2_P\ln({\hat\Omega}), \\
W&=& 2^{2k}\sqrt{\lambda_k}\, M^{3-2k}_P\,\bigg(\frac{1}{k} (H_u H_d)^k -\frac{2}{3(k+1)M^2_P}\,(H_u H_d)^{k+1} \bigg) \label{superpotential}
\eea
where $\lambda_k$ is the coupling parameter and $k$ is a positive integer.  We note that the coefficients of the two terms in the superpotential are fixed by the condition for the Higgs pole inflation and they can be maintained against loop corrections due to the non-renormalization theorem. 

As a result, the bosonic part of the Jordan-frame Lagrangian for $\phi_i=\{T, H_u, H_d\}$ in supergravity \cite{linde,jsugra} is given by
\bea
\frac{{\cal L}_J}{\sqrt{-g_J}}= -\frac{1}{2} M^2_P \,{\hat\Omega} \,R +|D_\mu H_u|^2 +|D_\mu H_d|^2 -V_J +3{\hat\Omega}  b^2_\mu
\label{sugraJ}
\eea
where the frame function and the F-term potential  are given by
\bea
{\hat\Omega} &=& T+{\bar T}- \frac{1}{3M^2_P}|H_u|^2- \frac{1}{3M^2_P}|H_d|^2, \\
b_\mu &=& -\frac{i}{2{\hat \Omega}} \Big(\partial_\mu\phi^i \partial_i {\hat\Omega}-\partial_\mu {\bar\phi}^{{\bar i}} \partial_{\bar i}{\hat\Omega}\Big),  \\
V_J&=&  {\hat V}_F +{\hat V}_D,
\eea
with
\bea
{\hat V}_F &=& \bigg|\frac{\partial W}{\partial H_u}\bigg|^2 +  \bigg|\frac{\partial W}{\partial H_d}\bigg|^2 \nonumber   \\
&=&2^{4k}\lambda_k M^{6-4k}_P (|H_u|^2+|H_d|^2) |H_u H_d|^{2(k-1)}\bigg|1-\frac{2}{3M^2_P}H_u H_d\bigg|^2,  \label{FtermJ} \\
{\hat V}_D &=&\frac{1}{8} g^{\prime 2} (|H_u|^2-|H_d|^2)^2 +\frac{1}{8} g^2 \Big((H_u)^\dagger{\vec \tau} H_u+ (H_d)^\dagger{\vec \tau} H_d\Big)^2.
\eea
Here, we note that there are extra kinetic terms for scalar fields from the auxiliary vector field $b_\mu$ in the Jordan frame. But, the cosmological evolution during inflation is dominated by the moduli $|\phi_i|$,  so we can take the scalar kinetic terms in the Jordan frame to be of canonical form \cite{linde,jsugra}. 
 
We also obtain the Einstein-frame Lagrangian in the following,
\bea
\frac{{\cal L}_E}{\sqrt{-g_E}}= -\frac{1}{2}  M^2_P R  + K_{i{\bar j}} \partial_\mu \phi^i \partial^\mu \phi^{\bar j} -V_E 
\eea 
where the Einstein-frame potential is 
\bea
V_E = V_F +V_D
\eea
with
\bea
V_F&=& e^{K/M^2_P} (D_i W K^{i{\bar j}} D_{\bar j} W^\dagger -3|W|^2/M^2_P), \\
V_D &=& \frac{{\hat V}_D}{{\hat\Omega}^2}. \label{DtermE}
\eea
Here,  the K\"ahler metric $K_{i{\bar j}}  =\partial_i \partial_{\bar j} K$ and the inverse K\"ahler metric $K^{i{\bar j}}$ with $i=T, H_u, H_d$ and ${\bar j}={\bar T}, H^\dagger_u, H^\dagger_d$ are given by
\bea
K_{i{\bar j}}&=&\frac{1}{{\hat\Omega}^2}\left(\begin{array}{ccc} 3M^2_P & -H_u  &  -H_d \\ -H^\dagger_u & T+{\bar T}- \frac{1}{3M^2_P}|H_d|^2& \frac{1}{3M^2_P}H^\dagger_u H_d  \\ -H^\dagger_d &  \frac{1}{3M^2_P}H_u H^\dagger_d  & T+{\bar T} - \frac{1}{3M^2_P}|H_u|^2\end{array}\right), \\
K^{i{\bar j}} &=&{\scriptsize{\hat\Omega}\left(\begin{array}{ccc} \frac{1}{3} (T+{\bar T}) & \frac{1}{3M^2_P}H_u & \frac{1}{3M^2_P}H_d \\  \frac{1}{3M^2_P}H^\dagger_u & 1 & 0  \\ \frac{1}{3M^2_P}H^\dagger_d & 0 & 1 \end{array}\right).}
\eea
As a result, we get the F-term potential in the Einstein frame as
\bea
V_F&=&  {\hat \Omega}^{-2}\bigg\{\bigg|\frac{\partial W}{\partial H_u}+\frac{1}{\hat \Omega} H^\dagger_u W\bigg|^2 +  \bigg|\frac{\partial W}{\partial H_d}+\frac{1}{\hat \Omega} H^\dagger_d W\bigg|^2 + \frac{3(T+{\bar T})|W|^2}{{\hat\Omega}^2}- \frac{3|W|^2}{{\hat\Omega}}  \nonumber \\
&&-\frac{1}{{\hat\Omega}}\bigg[H^\dagger_u W\bigg(\frac{\partial {\bar W}}{\partial H^\dagger_u}+\frac{1}{\hat \Omega} H_u{\bar W}\bigg)+H^\dagger_d W \bigg(\frac{\partial {\bar W}}{\partial H^\dagger_d}+\frac{1}{\hat \Omega} H_d {\bar W}\bigg)+{\rm h.c.} \bigg]\bigg\} \nonumber \\
&=& \frac{{\hat V}_F}{{\hat \Omega}^{2}}
\eea
where ${\hat V}_F$ is the F-term potential in Jordan frame, given in eq.~(\ref{FtermJ}). Thus, together with eq.~(\ref{DtermE}), the full scalar potential in Einstein frame is given by $V_E=({\hat V}_F+{\hat V}_D)/{\hat\Omega}^2$.

For the Higgs pole inflation, we take the inflaton direction along the CP-even neutral scalars, namely, $H_u=(0,h \sin\beta)/\sqrt{2}$ and  $H_u=(h \cos\beta,0)/\sqrt{2}$, so $|H_u|^2+|H_d|^2=\frac{1}{2}h^2$ and $H_u H_d=\frac{1}{2} h^2\sin\beta\cos\beta$. 
Then, taking the Higgs fields along the D-flat direction satisfying $\tan\beta=1$, we find it straightforward to see the consequences for the Higgs pole inflation in Jordan frame where the Jordan-frame potential in eq.~(\ref{sugraJ}) becomes
\bea
\frac{{\cal L}_J}{\sqrt{-g_J}}= -\frac{1}{2} M^2_P \bigg(T+{\bar T}-\frac{1}{6M^2_P}h^2 \bigg)R +\frac{1}{2} (\partial_\mu h)^2-V_J
\eea
with
\bea
V_J=8\lambda_k M^{6-4k}_P  h^{4k-2} \bigg(1-\frac{1}{6M^2_P}\,h^2 \bigg)^2.
\eea 
Therefore, after stabilizing the modulus at $T+{\bar T}=1$ by non-perturbative effects such as gaugino condensates in the hidden sector, we recover the non-supersymmetric Higgs pole inflation as discussed in the previous section. In this case, from the Einstein frame potential, $V_E=8\lambda_k M^{6-4k}_P  h^{4k-2}$, we find that the power of the Higgs potential is constrained to $h^{4k-2}$ with $k$ being the positive integer, namely, $h^2, h^6, h^{10}$, etc. 

In the supersymmetric case, the running Higgs quartic couplings coming from the D-term remain sizable because they are determined by the SM gauge couplings, so the gauge coupling unification ensures the vacuum stability. But, the Higgs fields are taken along the D-flat direction during inflation, so the effective Higgs quartic potential is dynamically relaxed to zero. In this case, there is no need to maintain individual running Higgs quartic couplings to be small, unlike the non-supersymmetric case. Then, the effective Higgs potential during inflation stems from the F-term whose value is fixed by a small coefficient  $\lambda_k$ of the superpotential term in eq.~(\ref{superpotential}). Therefore, the correct CMB normalization in eq.~(\ref{CMB}) is achievable. 

As for the non-supersymmetric case, the $k=1$ case is not compatible with the Higgs parameter required at low energy, so we need to take the power of the dominant term in the superpotential during inflation to satisfy $k\geq 2$. In this case, the $H_u H_d$ term in the superpotential must be suppressed. We can generate the effective $H_u H_d$ term by non-perturbative effects or the extra superpotential terms, such as $\Delta W=\lambda_S S H_u H_d$ or $\Delta W=\frac{\lambda_X}{\Lambda} X^2 H_u H_d$ in the superpotential \cite{muterm}, where $S, X$ being singlet chiral superfields frozen to zero during inflation.  In this case, we can reconcile the Higgs pole inflation with a realistic Higgs mass parameter in the context of supergravity.

\section{Conclusions}

We presented new results for the Higgs inflation based on the pole expansion of the effective Lagrangian. Both the effective Planck mass and the Higgs potential are in close proximity to zero during inflation, rendering the effective Higgs potential flat and allowing for a consistent inflationary prediction with the Planck data. Taking a single effective Higgs self-interaction term in the Einstein frame to be dominant, we found its coefficient and the rest of the effective interactions to be individually bounded by the CMB normalization. 

In the dominance with a Higgs quartic coupling, we need a very tiny Higgs quartic coupling during inflation,  which can be achieved by the renormalization group running of the Higgs quartic coupling towards the conformality. We also showed that higher order Higgs self-interactions with small coefficients can be kept for inflation while the Higgs quartic coupling is set to zero. This is possible in the supergravity embedding of the Higgs pole inflation where the Higgs inflaton is taken along the D-flat direction and the higher order Higgs self-interactions are separately obtained from the Higgs superpotential.

We have identified a class of Higgs pole models as the low-energy expansion near the conformal pole, showed testable predictions for inflation, and analyzed the perturbative reheating with a general equation of state as well as the preheating from an anharmonic Higgs potential. The presented models can be distinguishable from the original Higgs inflation and its sigma-model like models by the reheating processes.

\section*{Acknowledgments}

HML and AGM are supported in part by Basic Science Research Program through the National
Research Foundation of Korea (NRF) funded by the Ministry of Education, Science and
Technology (NRF-2022R1A2C2003567 and NRF-2021R1A4A2001897). 
SC thanks Yann Mambrini for his wise advice and Theoretical High Energy Physics group for the support for his stay during the Chung-Ang University BSM Workshop 2023. The authors acknowledge the support from FKPPL projects and from Institut Pascal at Universit\'e Paris-Saclay during the Paris-Saclay Astroparticle Symposium 2022, with the support of the P2IO Laboratory of Excellence (program "Investissements d'avenir" ANR-11-IDEX-0003-01 Paris-Saclay and ANR-10-LABX-0038), the P2I axis of the Graduate School of Physics of Universit\'e Paris-Saclay, as well as IJCLab, CEA, APPEC, IAS, OSUPS, and the IN2P3 master projet UCMN.




\end{document}